\documentclass[letterpaper, 10 pt, conference]{ieeeconf}  
\usepackage{amssymb}
\usepackage{amsmath}

\IEEEoverridecommandlockouts                              

\usepackage{graphics} 
\usepackage{epsfig} 

\usepackage{lipsum}
\usepackage{multicol}
\usepackage{amsmath} 

\usepackage{amsfonts}
\usepackage{physics}
\usepackage{algorithm}
\usepackage{algpseudocode}

\usepackage{subcaption}
\usepackage[font=footnotesize,labelfont=bf]{caption}

\usepackage{float}
\usepackage{cite}
\usepackage{color}

\usepackage{url}

\title{\LARGE \bf
Real-Time Formal Verification of Autonomous Systems With An FPGA
}

\author{Minh Bui, Michael Lu, Reza Hojabr, Mo Chen, Arrvindh Shriraman
\thanks{All the authors are with the School of Computing Science, Simon Fraser University. \{minh\_bui\_3, mla233, shojabro, mochen, ashriram\}@sfu.ca}
} 

\begin{document}
\maketitle
\thispagestyle{empty}
\pagestyle{empty}


\begin{abstract}
Hamilton-Jacobi reachability analysis is a powerful technique used to verify the safety of autonomous systems. This method is very good at handling non-linear system dynamics with disturbances and flexible set representations. A drawback to this approach is that it suffers from the curse of dimensionality, which prevents real-time deployment on safety-critical systems. In this paper, we show that a customized hardware design on a Field Programmable Gate Array (FPGA) could accelerate 4D grid-based Hamilton-Jacobi (HJ) reachability analysis up to 16 times compared to an optimized implementation and 142 times compared to MATLAB ToolboxLS on a 16-thread CPU. Our design can overcome the complex data access pattern while taking advantage of the parallel nature of the  HJ PDE computation. Because of this, we are able to achieve real-time formal verification with a 4D car model by re-solving the HJ PDE at a frequency of 5Hz on the FPGA as the environment changes. The latency of our computation is deterministic, which is crucial for safety-critical systems. Our approach presented here can be applied to different systems dynamics, and moreover, potentially leveraged for higher dimensions systems. We also demonstrate obstacle avoidance with a robot car in a changing environment. 
\end{abstract}


\section{Introduction}
\label{sec:intro}

Autonomous systems are becoming more prevalent in our lives. Examples of these systems include self-driving cars, unmanned aerial vehicles, rescue robots, etc. One key factor that will allow wider adoption of autonomous systems is the guaranteed safety of these systems. Despite tremendous progress in autonomous system research in areas such as motion planning, perception, and machine learning, deployment of these systems in environments that involve interactions with humans and other robots remains limited due to the potential danger these robotic systems can cause. Formal verification methods can help autonomous robots reach their untapped potential. 
 
Hamilton-Jacobi (HJ) reachability analysis is a formal verification approach that provides guaranteed safety and goal satisfactions to autonomous systems under adversarial disturbances. There are many ways to do reachability analysis, solving the HJ PDE is one way to characterize sets of safe states and synthesizes optimal controllers, which involves calculating Backward Reachable Tube (BRT) that describes a set of states the system must stay out of in order to avoid obstacles. HJ reachability analysis has been successfully applied in practical applications such as aircraft safe landing \cite{SafeLanding}, multi-vehicle path planning, multi-player reach avoid games \cite{SafePlatoon}. 
The appeal of this particular method is that it's very powerful in handling control and disturbances, nonlinear system dynamics, and flexible set representations.  

The main downside to HJ reachability is that it's solved on a multi-dimensional grid with the same number of dimensions as the number of state variables and scales exponentially with the number of dimensions. This prevents HJ formulation to be applied on real-time systems where safety is increasingly demanded. While 3D or smaller systems could be computed quickly with multi-core CPUs, practical systems that usually involve 4 to 5 system components can take several minutes to hours to compute. There have been researches that proposed decomposing high dimensional systems into smaller tractable sub-systems that can exactly compute \cite{ExacDec} or overapproximate the BRT in certain cases \cite{OverDec}. However, that challenge of applying HJ formulation on real-time systems remains, as some systems cannot be decomposed further than four dimensions, and over-approximation is introduced if projection methods are used. 

In this paper, we expand the limit of the number of dimensions for which we could directly compute the BRT in \emph{real time} through the use of FPGA. We would argue that customized hardware accelerators could complement well with those decomposition methods in making higher dimensional systems provably safe in real-time. As general-purpose computer no longer double its performance every two years due to the end of Moore's law, we have seen examples of successful hardware accelerations in other areas such as machine learning's training/inference \cite{TPU,Eyeriss, EIE}, robot's motion planning\cite{DukePaper}.

In this paper, our contributions are as follows: 
\begin{itemize}
    \item We prototype a customized hardware design on FPGA that accelerates HJ reachability analysis to 16x compared to state-of-the-art implementation and 142x compared to \cite{LsetToolbox1} on 16-thread CPU for 4D system
    \item We demonstrate that the system could meet real-time requirement of guaranteeing safety in changing environments by re-computing BRT at 5Hz 
    \item Demonstrate obstacle avoidance with a robot car driving in an environment in which new obstacles are introduced during run time at 5Hz.
\end{itemize}


 
\section{PRELIMINARIES}
\subsection{Hamilton-Jacobi Reachability Analysis}
Let $s \le 0$ be time and $z \in \mathbb{R}^n$ be the state of an autonomous system. The evolution of the system state over time is described by a system of ordinary differential equations (ODE) below.
\begin{equation}
\label{eq:main_ode}
    \dot{z} = \dv{z(s)}{s} = f(z(s), u(s), d(s))
\end{equation}
where $u(\cdot)$ and $d(\cdot)$ denote the control and disturbance function respectively. The system dynamics $f$ are assumed to be uniformly continuous, bounded and Lipschitz continuous in $z$ for fixed $u$ and $d$. Given $u(\cdot)$ and $d(\cdot)$, there exists a unique trajectory that solves equation \eqref{eq:main_ode}.

The trajectory or solution to equation \eqref{eq:main_ode} is denoted as $\zeta(s;z, t, u(\cdot), d(\cdot)): [t, 0] \rightarrow \mathbb{R}^n$ , which starts from state $z$ at time $t$ under control $u(\cdot)$ and disturbances $d(\cdot)$. $\zeta$ satisfies \eqref{eq:main_ode} almost everywhere with initial condition $\zeta(t;z, t, u(\cdot), d(\cdot)) = z$.

In reachability analysis, we begin with a system dynamics described by an ODE and a target set that represents unsafe states/obstacles \cite{HJOverview}. We then solve a HJ PDE to obtain Backward Reachable Tube (BRT), defined as follows:

\begin{equation}
\begin{aligned}
    \bar{\mathcal{A}} = \{z: \exists \gamma \in \Gamma, \forall u(\cdot) \in \mathbb{U}, \exists s \in [t, 0], \\ 	\zeta(s;z, t, u(\cdot), d(\cdot)) \in \mathcal{T} \}
\end{aligned}
\end{equation}

In HJ reachability analysis, a target set $\mathcal{T} \subseteq  \mathcal{R}^n$ is represented by the implicit surface function $V_{0}(z)$ as $\mathcal{T} = \{z: V_{0}(z) \leq 0\}$. The BRT is then the sub-level set of a value function $V(z, s)$ defined as below.
\begin{align}
\label{eq:eq3}
    V(z, s) = \min_{d(\cdot)}\max_{u(\cdot) \in \mathbb{U}}\min_{s\in[t,0]}V_0(\zeta(0;z, t, u(\cdot), d(\cdot)))
\end{align}
We assume disturbance is applied with non-anticipative strategies\cite{bansal2017hamilton}. In a zero-sum differential game, the control input and disturbances have opposite objectives.

The value function $V(z, s)$ can be obtained as the viscosity solution of this HJ PDE: 
\begin{equation}
\label{eq:HJ_variational_inequality}
\begin{gathered}
    \min\{D_{s}V(z,s) + H(z, \nabla V(z,s)), V(z,0) - V(z,s) \} = 0\\ 
    V(z,0) = l(z), s \in [t, 0] \\
    H(z, \nabla V(z, s)) =\min_{\gamma[u](\cdot)}\max_{u(\cdot)} \nabla V(z, s)^{T}f(z,u) \end{gathered}
\end{equation}

We compute the HJ PDE until it converges. Numerical toolboxes based on level set methods such as \cite{LsetToolbox1} are used to obtain a solution on a multi-dimensional grid for the above equation.

\subsection{Basic Numerical Solution}
Let us store the value function on a multi-dimensional grid, with the numerical solution of the value function denoted as $V$. Let $N_{d}$ be the grid size on the $d$th axis ($1 \le d \le 4$). We also let $x_{d,i}$ denote the state of grid $i$ in dimension $d$. 
In our approach throughout this paper, we will adopt the central differencing scheme for approximating derivatives in dimension $d$, which is defined as follows:
\begin{equation}
\label{eq:central_diff}
\begin{gathered}
    D_{d}^{-}V(x_{d, i}) = \dfrac{V(x_{d, i}) - V(x_{d, i-1})}{\Delta x_{d}}, \\
    D_{d}^{+}V(x_{d, i}) = \dfrac{V(x_{d, i+1}) - V(x_{d, i})}{\Delta x_{d}}, \\
    D_{d}V(x_{d, i}) = \dfrac{D_{d}^{+}V(x_{d, i}) + D_{d}^{-}V(x_{d, i})}{2} \end{gathered}
\end{equation}

The two terms $D_{d}^{-}$ and $D_{d}^{+}$ are the left and right approximations respectively. Note that for grid points at each end of each dimension (i.e $i = N_d-1$, $i = 0$), \eqref{eq:central_diff} is computed with extrapolated points.
The basic algorithm for solving \eqref{eq:HJ_variational_inequality} on-grid for 4D systems is then described as follows: 

\begin{algorithm}[h]
\caption{Value function solving procedures}
\label{HJalgorithm1}
\begin{algorithmic}[1]
  \State $V_{0}[N_1][N_2][N_3][N_4] \leftarrow l(z)$
    \State \texttt{//Compute Hamiltonian term, and max, min deriv}
  \For{\texttt{$i = 0 :N_1 - 1$; $j= 0:N_2 -1$; $k= 0:N_3 -1$; $l= 0:N_4 -1$}}   
        \State \texttt{Compute \eqref{eq:central_diff} for } $1 \leq d \leq 4 $ 
        \State $ minDeriv \leftarrow min(minDeriv, D_{d}V(x))$
        \State $ maxDeriv \leftarrow max(maxDeriv, D_{d}V(x))$
        \State  $\displaystyle u_{opt} \leftarrow \arg \max_{u \in \mathbb{U}} \nabla V(z, s)^{\top}f(z,u)$  
        \State $\dot{x} \leftarrow f(z, u_{opt})$
        \State $H_{i, j, k, l} \leftarrow \nabla V(z, s)^{\top}\dot{x}$
        
  \EndFor
  \State \texttt{// Compute dissipation and add to H}
  \For{\texttt{$i = 0 :N_1 - 1$; $j= 0:N_2 -1$; $k= 0:N_3 -1$; $l= 0:N_4 -1$}}
  \State $\alpha_d(x) \leftarrow \max_{p \in [minDeriv, maxDeriv]} \abs{\dfrac{\partial H(x,p)}{\partial p_d}}$
  \State $H_{i, j, k, l} \leftarrow H_{i, j, k, l} - \Sigma_{d=1}^{4} \alpha_{d}(x)\dfrac{D_{d}^{+}\check(x) - D_{d}^{-}\check(x) }{2}$
  \State $\alpha_{d}^{max} \leftarrow max(\alpha_{d}^{max}, \alpha_{d})$
    \EndFor
\State \texttt{//Compute stable integration time step}
\State $\Delta t \leftarrow (\Sigma_{d=1}^{4}\dfrac{\abs{\alpha_{d}^{max}}}{\Delta x_{d}})^{-1}$
  \State $V_{t+1} \leftarrow H\Delta{t} + V_{t}$
  \State \texttt{$V_{t+1} \leftarrow min(V_0, V_{t+1})$}
  \State \texttt{$\epsilon \leftarrow \abs{V_{t+1} - V_{t}}$}
  \If{$\epsilon < threshold$}
    \State \texttt{$V_{t} \leftarrow V_{t+1}$}
    \State \texttt{Go to line 3}  
  \EndIf
\end{algorithmic}
\end{algorithm}
The above algorithm loops through the 4D array three times. In the first grid iteration, the Hamiltonian terms, maximum and minimum derivative is determined (lines 3-9). In the next grid iteration, the dissipation is computed and added to the Hamiltonian in order to make the computation stable. At the same time, the maximum alphas in all dimensions defined in line 13 are computed. These $\alpha_d^{max}$ are used to determine the step bound $\Delta t$. In the third grid iteration (line 19), each grid point is integrated for a length of $\Delta t$.
\begin{figure}[H]
    \centering
    \includegraphics[width=.30\textwidth]{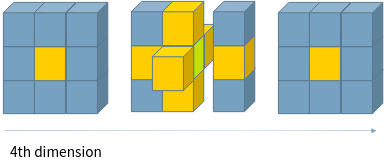}
    \caption{9 memory accesses (yellow + green colored grid) within each iteration for computing derivatives in all 4 directions as in line 3 (algorithm \ref{HJalgorithm2})}
    \label{fig:my_label}
\end{figure}
In certain cases, $\alpha_{d}(x)$ in line 13 is the same as computing the absolute value of $\dot{x}$, which has been computed in line 8. In addition, in a lot of cases, $\alpha_{d}^{max}$ stays the same across different time iterations. We also observed that $\Delta t$ depends only on grid configuration and $\alpha_{d}^{max}$. So instead of re-computing $\Delta t$ every time and then loop through the 4D grid array again, we could pre-compute $\Delta t$ and re-use it for all the time iterations. Combining these ideas together, throughout this paper, we will use the following algorithm with one grid looping, which is more computationally efficient:
\begin{algorithm}[h]
\caption{Value function solving procedures}
\label{HJalgorithm2}
\begin{algorithmic}[1]
  \State $V_{0}[N_1][N_2][N_3][N_4] \leftarrow l(z)$
  \For{\texttt{$i = 0 :N_1 - 1$; $j= 0:N_2 -1$; $k= 0:N_3 -1$; $l= 0:N_4 -1$}}
        \State \texttt{Compute \eqref{eq:central_diff} for } $1 \leq d \leq 4 $ 
        
        \State  $\displaystyle u_{opt} \leftarrow \arg \max_{u \in \mathbb{U}} \nabla V(z, s)^{\top}f(z,u)$  
        \State $\dot{x} \leftarrow f(z, u_{opt})$
        \State $H_{i, j, k, l} \leftarrow \nabla V(z, s)^{\top}\dot{x}$
        \State $H_{i, j, k, l} \leftarrow H_{i, j, k, l} - \Sigma_{d=1}^{4} \abs{\dot{x_{d}}}\dfrac{D_{d}^{+}(x) - D_{d}^{-}(x) }{2}$
        \State $V_{t+1,(i, j, k, l)} \leftarrow H_{i, j, k ,l}\Delta{t}_{precomputed} + V_{t, (i, j, k, l)}$
        \State \texttt{$V_{t+1, (i, j, k, l)} \leftarrow min(V_{0, (i, j, k, l)}, V_{t+1, (i, j, k, l)})$}
  \EndFor
  \State \texttt{$\epsilon \leftarrow \abs{V_{t+1} - V_{t}}$}
  \If{$\epsilon < threshold$}
    \State \texttt{$V_{t} \leftarrow V_{t+1}$}
    \State \texttt{Go to line 2}  
  \EndIf
\end{algorithmic}
\end{algorithm}
\subsection{Field Programmable Gate Arrays (FPGA)}
FPGA are configurable intergrated circuits that are programmed for specific applications using hardware description language (HDL). 

\begin{figure*}[h]
  \includegraphics[width=\textwidth]{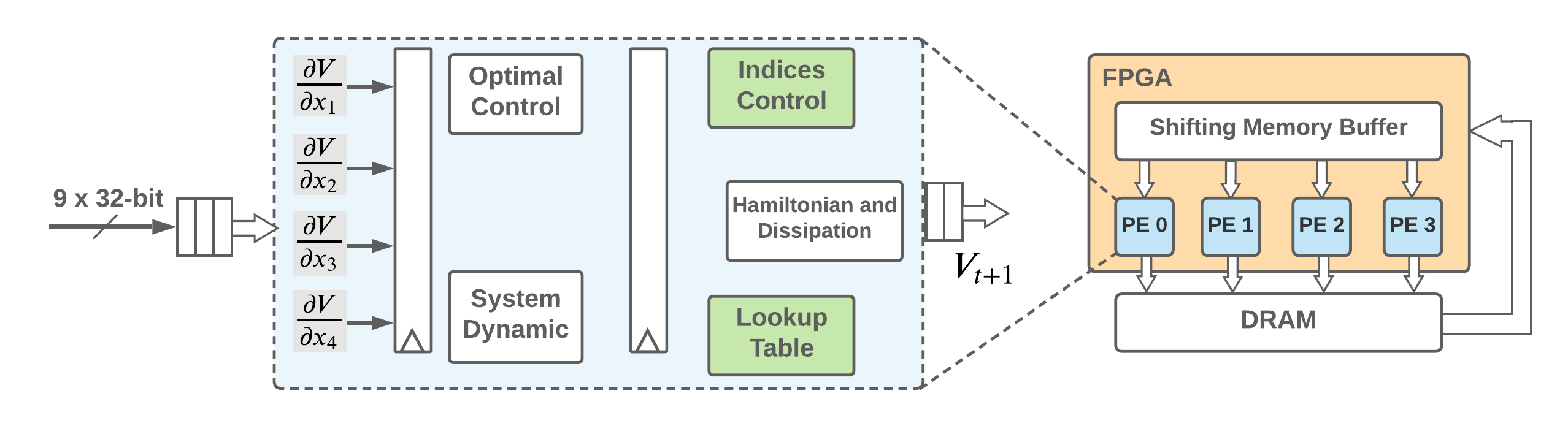}
  \caption{System overview on FPGA (Right). The initial value array is first transferred from DRAM to FPGA's on-chip memory. The memory buffer then distributes data to the 4 PEs to concurrently compute the new value function at 4 consecutive grid points. The output from PE is then written back to DRAM. Each fully pipelined PE outputs one grid point every clock cycle (Left). Inside the PE, there are hardware components that sequentially solve algorithm 2}
    \label{fig:overall_system}
\end{figure*}

Computing platforms such as CPUs, GPUs, and FPGAs have a memory component and computing cores. Compute cores must request and receive all the necessary data from the memory component before proceeding with the computation. If the memory component cannot provide all data accesses the application requires to proceed at once, cores have to stall and wait, slowing down the computation.
\begin{figure*}[h]
  \includegraphics[width=\textwidth, height=155.5pt]{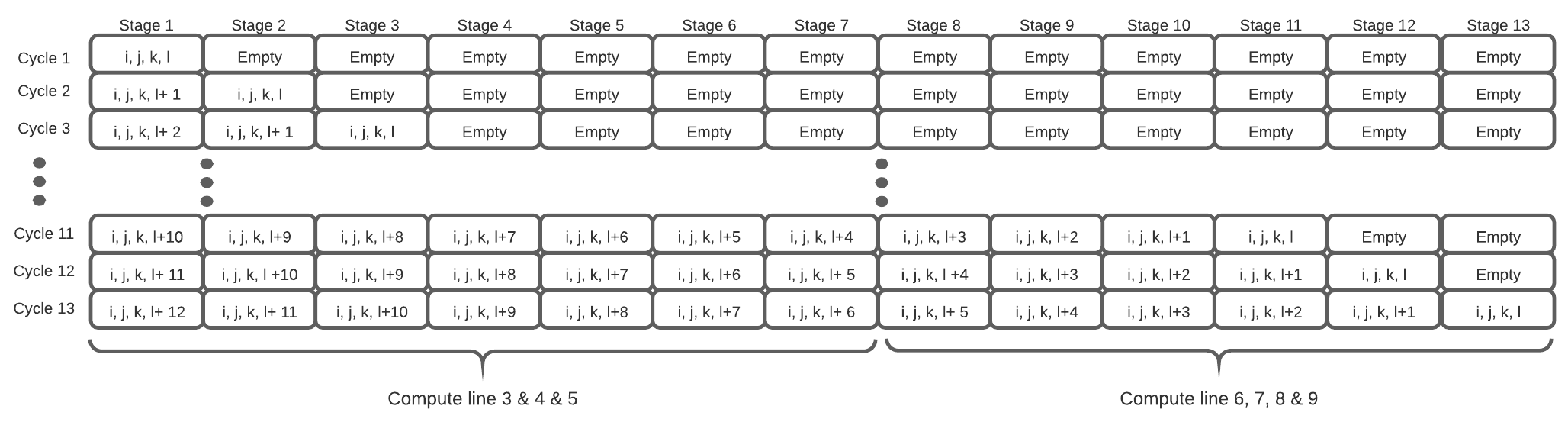}
  \caption{Pipelining schedule of a single PE. The PE's operation is an assembly line where multiple grid points could be processed at the same time at different stages. Each stage is physical hardware that computes specific parts of algorithm 2. At a particular stage and a particular cycle, the PE is busy computing a certain part of algorithm 2 for the grid point at the indices shown. Note that for simplicity, the indices shown here are for a single PE only.}
  \label{fig: pipelining_schedule}
\end{figure*}
Efficient systems need to have both fast computing cores and fast data distributions from the memory. Depending on the application, the memory access and computing pattern will vary. General-purpose CPU/GPU are often architected towards a reasonable performance for a wide variety of applications, but unoptimized for any particular application. FPGA chip, on the other hand, provides programmable digital circuits to design customized computing core and memory blocks. Thus, one can leverage knowledge about the details of the computing workload to design an efficient system accordingly with FGPA. With FPGA, one could control and achieve a higher degree of parallelism from the digital hardware level at the cost of programmability.


\subsection{Problem Description}

A key observation of algorithm 2 is that each new grid point $V_{t+1}$ could be computed independently with each other within one time iteration and therefore, in parallel. We could then leverage a high degree of parallelism on FPGA by having many cores to update as many grid points concurrently as possible. 

However, before that, two challenges must be addressed. Firstly, memory blocks need to efficiently distribute data to compute cores. In order for a loop computation to proceed, each of these cores needs up to 9 data inputs (Fig 2.) and a memory design needs to satisfy this.  
Secondly, a four-dimensional grid takes up tens of megabytes in memory and therefore cannot be fully fit to FPGA's on-chip memory for fast access. 


In this paper, our goal is twofold. First, we will discuss our hardware design that can solve the above challenges and maximize parallel computation of algorithm 2 while efficiently making use of FPGA's on-chip memory. Next, we will show that this enables low latency of computation on FPGA which could be deployed in real-time systems.


\section{Solving the HJ PDE Using FPGAs}
\label{sec:fpga}

Before going into details of the design, we will introduce some terminologies that will be relevant throughout the next section.

In digital systems, time is discretized into the unit of a \emph{clock cycle}, which is the amount of time it takes for an operation such as computing, loading, and storing to proceed. Each clock cycle typically is a few nanoseconds. 
Dynamic Random Access Memory (DRAM) is a type of memory that sits outside of the FPGA, which has higher memory capacity but takes a lot more clock cycles to access.

Our custom hardware comprised two main components: on-chip memory buffer, and processing elements (PE) or computing cores (shown in Fig \ref{fig:overall_system}). The memory buffer is on-chip storage, providing access to all the grid points a PE needs to compute a new value function. Each PE is a digital circuit that takes 9 grid points from the memory buffer to compute a new value function at a particular grid point according to algorithm 2 (line 3-10). In the following subsections, we will go into the details of each component.
\subsection{Indexed Processing Element (PE)}

The PE has the following target design objectives: 
(1) increase compute throughput (defined as the number of output generated per second) through pipelining, (2) reduce the computation time of each PE, (3) and ensure the correctness of result while minimizing data transfer between DRAM and FPGA.

In our design, we use 4 PEs (as shown in figure \ref{fig:overall_system}). Each PE has an index $idx$ with $0 \le idx \le 3$ associated with it and computes the grid point $V_{t+1}(i, j, k, l + idx)$. At the beginning of the computation of algorithm 2, each PE takes as input a grid index $(i,j,k,l)$ and its 8 neighbours to start computing $V_{t+1}(i,j,k,l)$ according to algorithm 2 (line 2-10).

To increase computation throughput, each PE is fully pipelined. Similar to an assembly line, the PE operation is divided into multiple stages taking a few clock cycles to complete (Fig. \ref{fig: pipelining_schedule}). Each stage within the pipeline is physical hardware that has a computation corresponding to one of the lines in algorithm 2 (line 3-10) for a particular index $i, j, k, l$. Every clock cycle, the result from previous stages will be loaded to the next stage, following the sequential order of algorithm 2. At any time during operations, the processing element is computing different intermediate components for multiple indices concurrently (explained in Fig.\ref{fig: pipelining_schedule}). 

To ensure that the computation is correct, inside each of the PE, there are indices counters to keep track of loop variable $i, j, k, l$, with the inner loop variable incrementing by one every clock cycle. These indices are used to correctly address the state vectors during system dynamics computation. To avoid accessing external DRAM we store these 4 state/position vectors $x$ or any fixed non-linear functions such as $\cos(\cdot)$ and $\sin(\cdot)$ of these states as a lookup table stored in on-chip memory, as state vectors depend only on grid configuration and do not change with the environment. Each PE will have its own look-up table to avoid communications between PEs. Having this data on-chip will only require a few kilobytes of memory and no need to access DRAM throughout the computation. 
\begin{figure*}[h]
  \centering
  \includegraphics[width=0.85\textwidth, height=250pt]{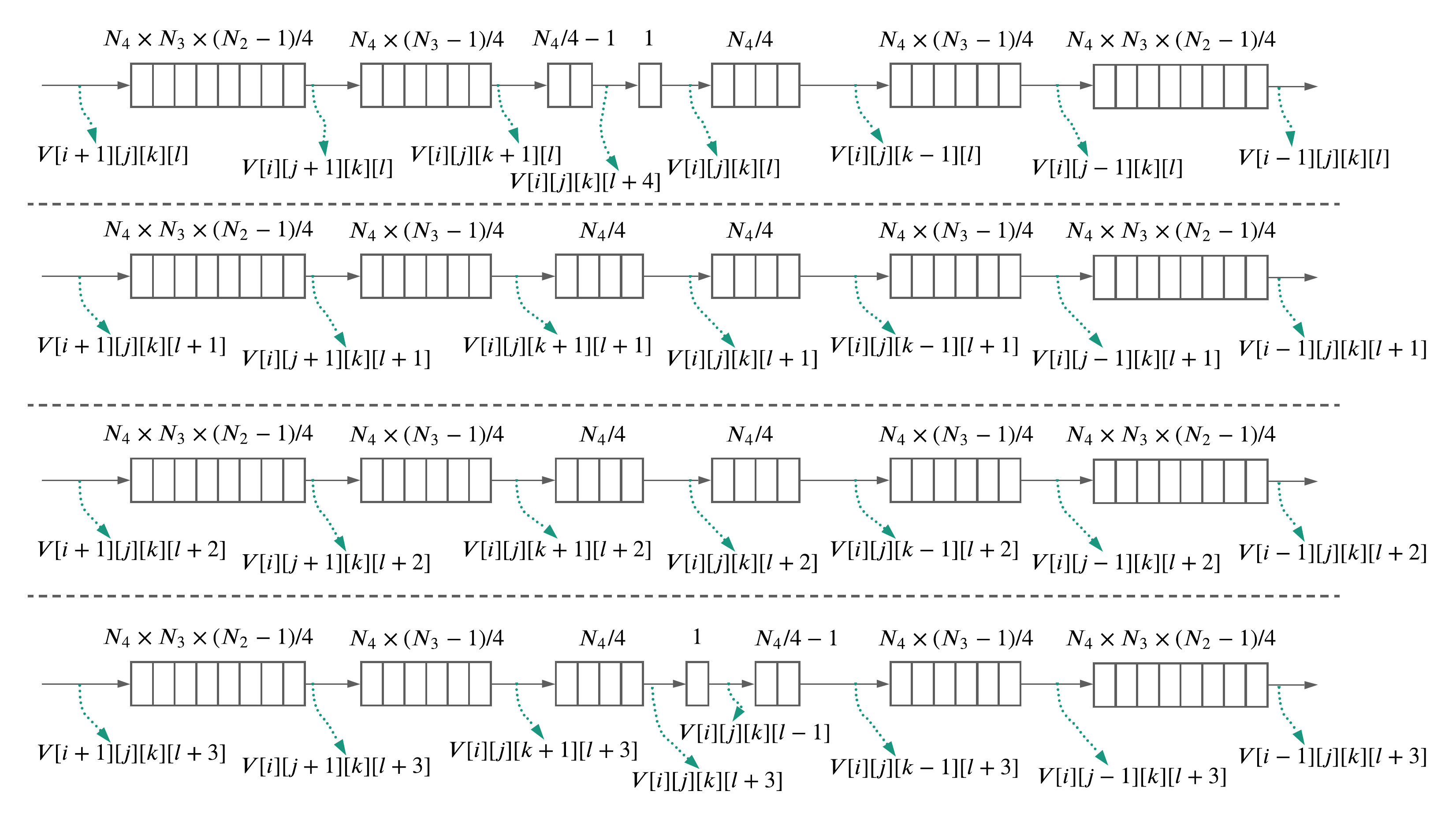}
  \caption{Four lines of memory buffer supply all the grid data to the four PEs. Each of the rectangle blocks is a FIFO queue synthesized as Block RAM (BRAM). The overhead notation is the size of the FIFO queue with $N_1, N_2, N_3, N_4$ as the four grid dimensions. Note that the queue's size depends only on three dimensions. Every clock cycle, new grid points streamed from DRAM start entering each buffer line (left-hand side) and grid points at the end of the lines are discarded (right-hand side). }
  \label{fig: parallel_line}
\end{figure*}
\subsection{On-Chip Memory Buffer}
The memory buffer has the following key design objectives: (1) minimizing the amount of on-chip memory usage and external DRAM accesses while (2) concurrently providing 9 grid points to each PE every clock cycle.

One problem of working with a high-dimensional grid is that the whole grid can take up tens of megabytes and therefore cannot be fully fit to a state-of-the-art FPGA's on-chip memory. Instead of storing everything on-chip, in our design, grid points are streamed continuously from DRAM into an on-chip memory buffer (shown in Fig.\ref{fig:overall_system}) and can be re-used many times for spatial derivatives computation in 4 dimensions before being thrown away. From the grid dimensions, we could compute the maximum reuse distance beyond which a grid point can be safely discarded as no longer needed. This maximum reuse distance is equal to the minimum size of on-chip memory buffer, which is dependent only on $N-1$ dimensions \cite{soda} and can be fitted to an FPGA's on-chip memory. Our memory buffer structure is implemented as First In First Out (FIFO) queue data structure. Every clock cycle, a new grid point supplied from DRAM will start entering the FIFO queue while the grid point reaching at the end of the FIFO queue will be discarded (shown in Fig. \ref{fig: parallel_line}).

FPGA on-chip memory buffers are composed of standard Blocks of Random Access Memory (BRAM). Each BRAM has two-ports and at most two reads can be requested concurrently in the same clock cycle. If all 9 grid points (shown in Fig. \ref{fig:my_label}) are stored in the same BRAM at the same time, a PE would then have to wait for 5 clock cycles before performing the computation. One way to increase the number of accesses per clock cycle is to duplicate the data in multiple BRAMs, but this would not work well for multidimensional arrays since these array copies easily exceed FPGA on-chip memory. A different technique would be \emph{memory banking}, which is to partition the memory on-chip into multiple BRAM that could concurrently deliver data to the PE, allowing the PE to start to compute new value function for a grid point in one clock cycle. 

To allow concurrent access for multiple PEs, we adopted the parallel memory buffer microarchitecture from \cite{soda}. Corresponding to the number of PEs, our on-chip storage structure is made of 4 line buffers. Each of these line buffers is a sequence of BRAM connected acting in a queue fashion: a grid point moves towards the end of the line every clock cycle. The two endpoints of each BRAM (shown in Fig 4) provide tapping points that are connected as inputs to the PEs.
The number of PEs, therefore, is mainly limited by the DRAM bandwidth. 

We also made modifications to the execution flow in \cite{OptimalMicro} to accommodate for computing values function at the boundary. Once each of the buffer lines is half full, all the processing elements can start computing a new value function.    

\subsection{Fixed-Point Representation}
Computing a new value function based on algorithm 2 involves multiple addition operations on floating-point numbers. At the hardware level, the addition of floating-point numbers is as computationally expensive as fixed-point multiplication, which would take up lots of resources and chip's area. Instead, we use fixed-point representations for our data to reduce the burden on the hardware. We will show in the next section that this has little impact on the correctness of the computation if the radix point is chosen carefully for the grid configuration.    

\section{EXPERIMENT \& RESULT}
\label{sec:exp}
\subsection{Experiment setup}
In this section, we demonstrate that our system can meet the real-time requirement through an obstacle avoidance demonstration in a changing environment.

We used a Tamiya TT02 model RC car\cite{nvidiajetracer} controlled by an on-board Nvidia Jetson Nano microcontroller inside a $4$m $\times$ $6$m room. We use the following extended Dubins car model for its dynamics:
\begin{equation}
\begin{split}
\label{eq:car_dynamics}
    \dot{x} &= v \cos(\theta)\\
    \dot{y} &= v \sin(\theta)\\
    \dot{v} &=a\\
    \dot{\theta} &= v\frac{\tan(\delta)}{L}
\end{split}
\end{equation}

\noindent where $a \in \left[-1.5, 1.5\right]$, 
$\delta \in \left[-\frac{\pi}{18}, \frac{\pi}{18}\right]$, 
and $L = 0.3$m. The control inputs are the acceleration $a$ and the steering angle $\delta$. We use a grid size of $60\times60\times20\times36$ with resolutions of $0.1$m, $0.067$m, $0.2$m/s and $0.17$ rad for $x$-position, $y$-position, speed and angle, respectively.

Inside the room, we use orange cones as obstacles and a motion capture system is used to accurately track the car's state and the position of the obstacles. We initialize the initial value function as follows:
\begin{equation}
\label{eq:init_V}
    V_{0}(x, y, v, \theta) = \sqrt{(x-x_o)^2 + (y-y_o)^2} - R
\end{equation}
where $x_o$ and $y_o$ are the obstacle's positions and R is the radius of the cone.
Obstacle's positions can be obtained from the motion capture system. Each of the cones has a radius of $0.08$m but is set as $0.75$m to account for the model mismatch between the car and the dynamics used.

\begin{figure}[h]
\centering
\begin{subfigure}[t]{.4\textwidth}
\centering
\includegraphics[width=\linewidth]{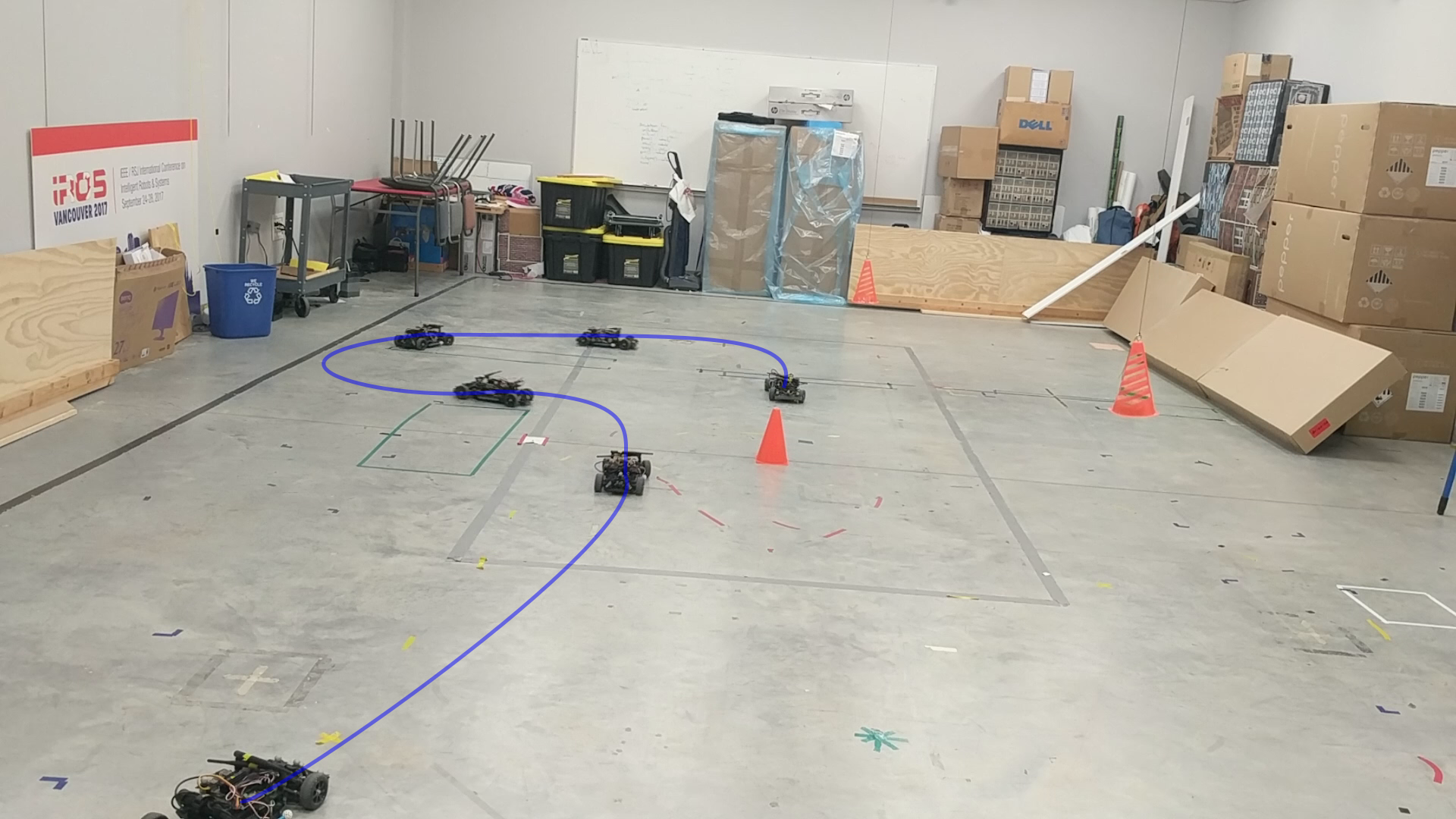}
\caption{Environment 1}
\end{subfigure}
\hfill
\begin{subfigure}[t]{.4\textwidth}
\centering
\includegraphics[width=\linewidth]{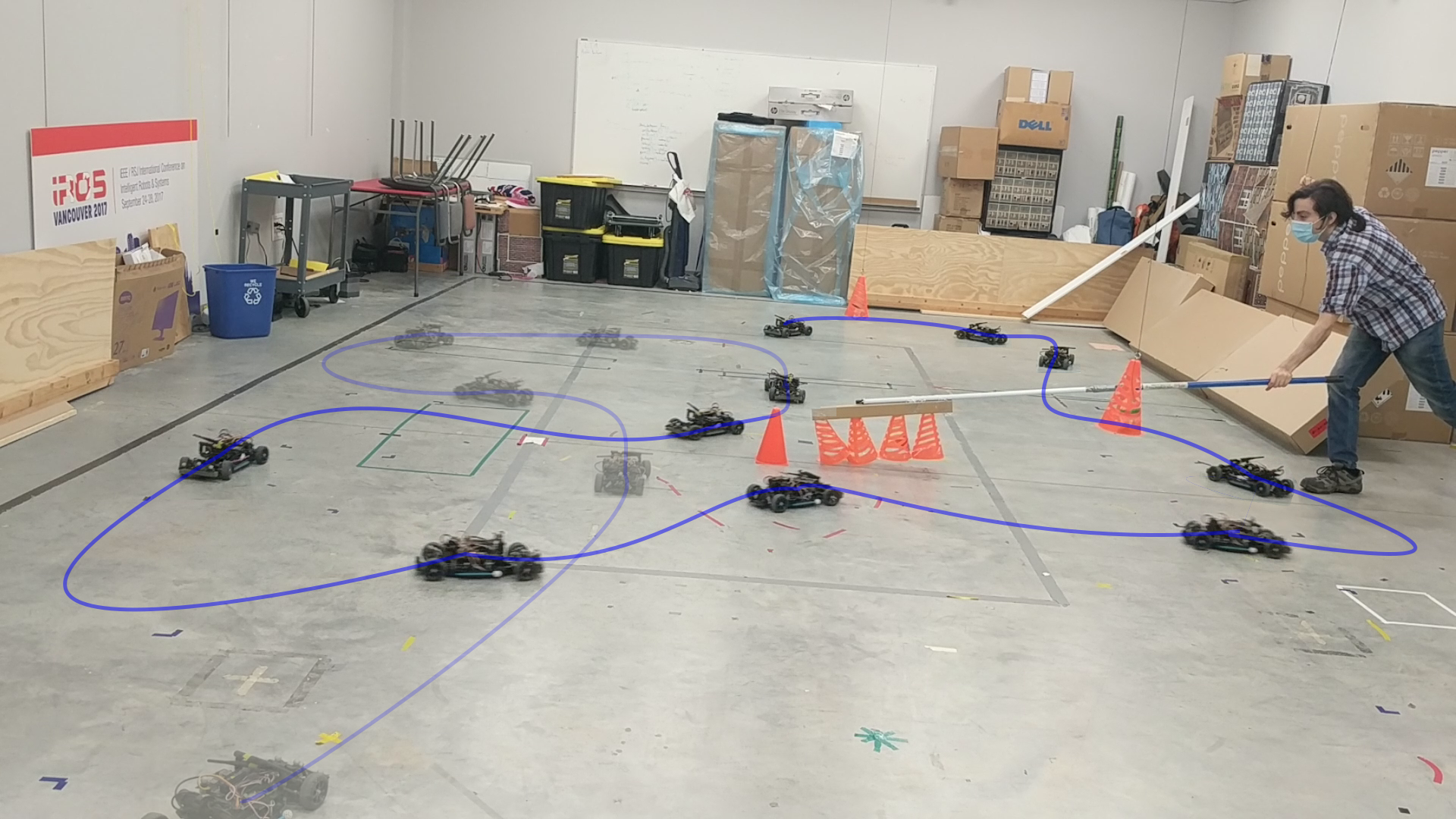}
\caption{Environment 2}
\end{subfigure}
\hfill
\begin{subfigure}[t]{.4\textwidth}
\centering
\includegraphics[width=\linewidth]{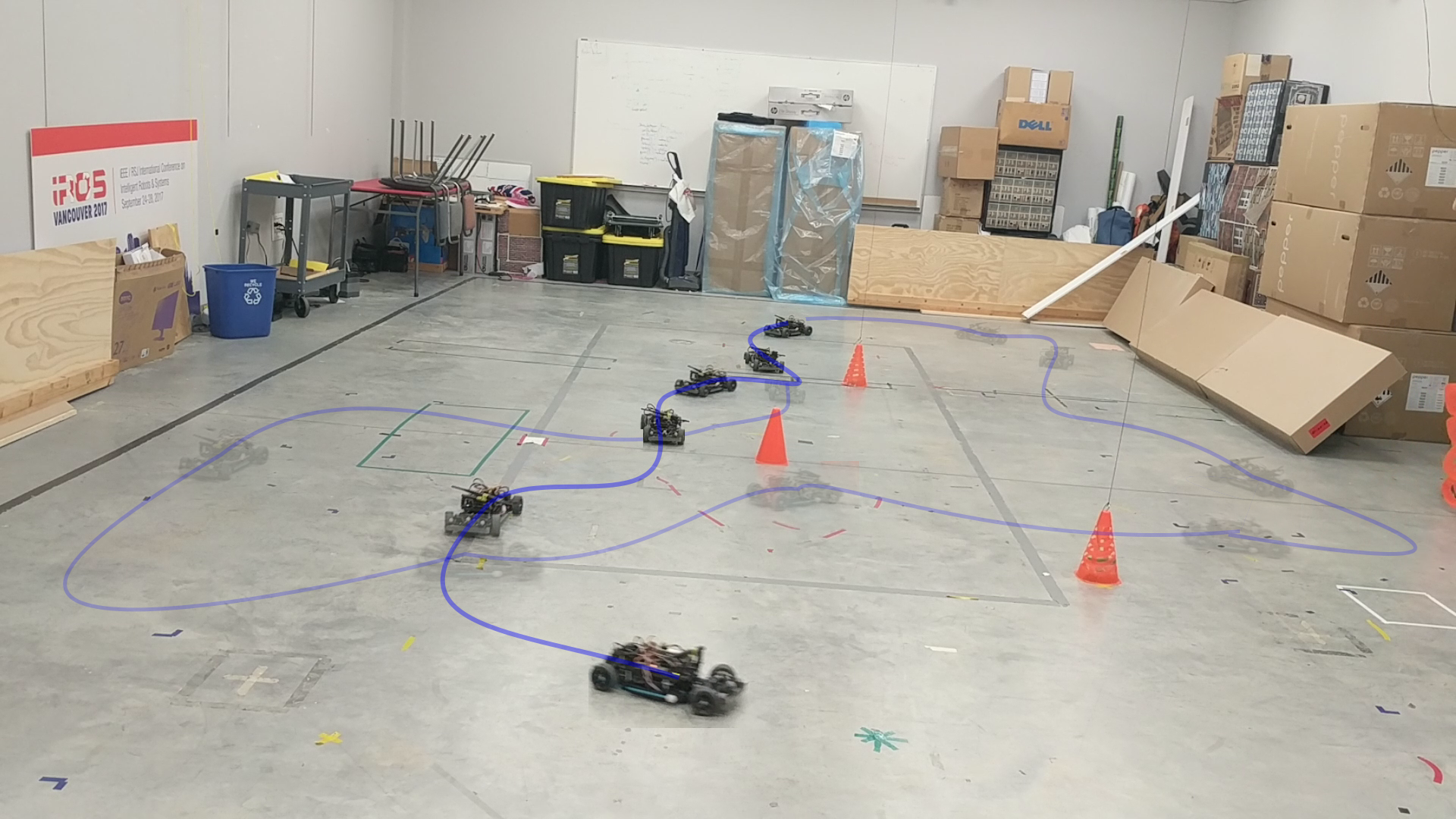}
\caption{Environment 3}
\end{subfigure}
\caption{Different BRTs are used as the placement of cones change over time which limits where the RC car can be as it drives around the room.}
\label{fig:cones}
\end{figure}

For the experiment, we considered three different environments, with different cone placements, set up inside the room as shown in Fig. \ref{fig:cones}. For each environment, a user manually controls the car and tries to steer into the cones.
\begin{equation}
\label{eq:closetoboundary}
    V(x,y,v, \theta) < 0.15
\end{equation}
Given the car's state, when \eqref{eq:closetoboundary} is satisfied, the car is near the boundary of a BRT so optimal control computed from the value function is applied to safely avoid the cone. The optimal control is obtained from the value function as follows:
\begin{equation}
\label{eq:uopt_max}
    u_{opt} =  \arg \max_{u \in \mathbb{U}} \nabla V(x, y, v, \theta , s)^{\top}f(x, y, v, \theta, u)
\end{equation}

We pre-compute the BRTs with a horizontal time of $0.5$s for three environments using optimized\_dp\cite{optimizedp} and demonstrate safety by correctly loading the value functions as the environment changes. 
We choose to pre-compute the BRTs in order to emulate having an FPGA on-board without extra latency resulted from communication with a remote AWS instance.
For all environments, the maximum time step to make the integration stable is $0.007497$s. Initially, the room had a single cone but changed over time to different cone placements. 
The BRT of a new environment could not be used until 200ms after the environment has changed, which is longer than the time taken to compute the same BRT on an FPGA.  A video of these experiments can be found at \url{https://www.youtube.com/playlist?list=PLUBop1d3Zm2vgPL4Hxtz8JufnIPmvrwlC}

\subsection{Hardware Correctness}
We use fixed-point data representations for hardware computation. In particular, we use 32 bits with 5 bits to represent the integer part (including sign) and 27 bits for the decimal part. With this choice, the precision of our computation is $2^{-27}=7.45\times10^{-9}$ and the range of our computation is from $-16$ to $16$. The area we use for the experiment is $4$m $\times$ $6$m, hence the largest absolute distance is the diagonal of $7.2$m. Therefore, the number of integer bits is enough to represent all possible values in the solution $V$, which has the physical interpretation of minimum distance to collision over time, given \eqref{eq:eq3} and the choice of $V_0$ in \eqref{eq:init_V}.

We choose to synthesize and implement our design on AWS F1 FPGA because of its flexibility and availability. To correctly input data to the FPGA, we first generate an initial value array based on the obstacles' positions and radius described by \eqref{eq:init_V}. Then this value array is converted from floating-point to fixed-point number correctly based on the bit choice discussed above. Afterward, the value array is passed to the FPGA for the HJ PDE solving procedure to start.

For all three experiment, we verified the correctness of BRT generated by our hardware with the toolbox at \cite{optimizedp} by comparing the maximum error between corresponding grid points. The toolbox uses 32-bit floating-point numbers. The numerical error resulting from the different representations is shown in table below for the three environments in table \ref{table:Error}.
\begin{center}
\captionof{table}{ERROR COMPARISON} 
 \label{table:Error}
 \begin{tabular}{||c c c c||} 
 \hline
  & Env. 1 & Env. 2 & Env. 3 \\ [0.5ex] 
 \hline\hline
 Error & $1.68\times10^{-6}$ & $1.78\times10^{-6}$& $1.37\times10^{-6}$ \\ 
 \hline
\end{tabular}
\end{center}
These negligible errors are due to precision difference between fixed-point and floating point number. Even though the computation is repeated for many iterations, the maximum error does not grow dramatically over time. We believe that is because of the convergence property of BRT. As time grows, the rate of changes in the grid values also slows down leading to stable discrepancy between the correct floating point and fixed-point values.

\subsection{Computational speed and Resources Usage}
To measure the speed up for all three environments, we compare the computation time on AWS FPGA running at 250MHz against \cite{optimizedp} and \cite{LsetToolbox1} running on a 16-thread Intel(R) Core(TM) i9-9900K CPU at 3.60GHz. The latency here is the time it takes to compute the BRT. For FPGA, latency can be computed by multiplying the clock cycles with the clock period. The result is summarized in the table below.

\begin{center}
\captionof{table}{FPGA}\label{table:fpga} 
 \begin{tabular}{||c c c c c||} 
 \hline
    & \ \textbf{Clock cycles} & \ \textbf{Period} & \textbf{Iterations} & \textbf{Latency} \\ [0.4ex] 
 \hline\hline
 Env. 1 & 44155209 & 4 ns & 67 & 0.176s\\ 
 \hline
 Env. 2 & 44155209 & 4 ns & 67 & 0.176s\\
 \hline
 Env. 3 & 44155209 & 4 ns & 67 & 0.176s\\ [0.5ex]
 \hline
\end{tabular}
\end{center}

\begin{center}
\captionof{table}{optimized\_dp\cite{optimizedp}} \label{table:op_dp} 
 \begin{tabular}{||c c c c||} 
 \hline
    & \ \textbf{Latency} & \textbf{Iterations} & \textbf{FGPA speed up} \\ [0.1ex] 
 \hline\hline
 Env. 1 & 3.35 s & 67 & \textbf{$\times$18.9} \\ 
 \hline
 Env. 2 & 2.99 s & 67 & \textbf{$\times$17} \\
 \hline
 Env. 3 & 3.42 s & 67 & \textbf{$\times$19.4}\\ [0.5ex]
 \hline
\end{tabular}
\end{center}

\begin{center}
 \captionof{table}{ToolboxLS\cite{LsetToolbox1}} \label{table:toolboxls} 
 \begin{tabular}{||c c c c||} 
 \hline
    & \ \textbf{Latency} & \textbf{Iterations} & \textbf{FPGA speed up} \\ [0.1ex] 
 \hline\hline
 Env. 1 & 25.11 s & 70 & \textbf{$\times$142} \\ 
 \hline
 Env. 2 & 25.14 s & 70 & \textbf{$\times$142} \\
 \hline
 Env. 3 & 25.18 s & 70 & \textbf{$\times$142}\\ [0.5ex]
 \hline
\end{tabular}
\end{center}
 
It can be observed that the latency of computation on FPGA is fixed and deterministic for all three environments while the latency on CPUs varies even though the computation remains the same. With the lower latency of $0.176$s, we are able to update the value function at a frequency of $5.68$Hz. The resources usage of our design for 4 PEs is shown in the table below. 
\begin{center}
 \begin{tabular}{||c c c c||} 
 \hline
    & \ \textbf{LUT} & \textbf{BRAM} & \textbf{DSP} \\ [0.1ex] 
 \hline\hline
 \textbf{Used} & 26319 & 519 &  598\\ 
 \hline
 \hline
 \textbf{Available} & 111900 & 1680 & 5640\\ 
 \hline
 \textbf{Utilization} & \textbf{14.03\%} & \textbf{30.89\%} & \textbf{10.6\%} \\[0.5ex]
 \hline
\end{tabular}
\end{center}
On an FPGA, arithmetic operations on numbers are implemented using Digital Signal Processing (DSP) hardware or Look Up Table (LUT) that perform logical functions. Our design does not significantly consume most of the available resources and could be scaled up to a larger grid size.

\section{CONCLUSION}
This paper introduces a novel customized hardware design on FPGA that allows HJ reachability analysis to be computed $16$x faster than state-of-the-art implementation on a 16-thread CPU. Because of that, we are able to solve the HJ PDE at a frequency of 5Hz. The latency of our computation on FPGA is deterministic for all computation iterations, which is crucial for safety-critical systems. Our design approach presented here can be applied to system dynamics and potentially higher dimensional systems. 
Finally, we demonstrate that at 5Hz, a robot car can safely avoid obstacles and guarantee safety.

\addtolength{\textheight}{-12cm}   








\bibliographystyle{ieeetr}
\bibliography{references.bib}

\end{document}